\documentclass{PoS}

\title{Recent Results from RENO}

\ShortTitle{Recent Results from RENO}

\author{\speaker{Myoung Youl Pac}\thanks{This work was supported by NRF Grant No 2009-0083526 and 2016R1D1A3B02010606. RENO gratefully acknowledge the cooperation of the
Hanbit Nuclear Power Site and the Korea Hydro \& Nuclear Power Co., Ltd.
(KHNP) and also thans to KISTI  for providing computing and network resources
through GSDC, and all the technical and administrative people who greatly
helped in making this experiment possible. The Author wishs to acknowledge Dongshin University for supporting my applicication for a sabbatical and also thanks KIAS for helpful discussion and support for my staying in there.}, {on behalf of RENO Collaboration}\\
        Dongshin University, Naju 58245, Korea\\
        E-mail: \email{pac@dsu.ac.kr}}

\abstract{RENO (Reactor Experiment for Neutrino Oscillation) is the reactor neutrino experiment which
has been taking data from August 2011 with two identical near and far detectors at Hanbit Nuclear Power Plant, Yonggwang, Korea. Using 1,500 live days
of data, $\sin^2 2\theta_{13}$ and $|\Delta m^2 _{ee}|$ are updated using spectral measurements: $\sin^2 2\theta_{13} = 0.086 \pm 0.006 ({\rm stat.}) \pm 0.005
({\rm syst.})$ and $|\Delta m^2 _{ee}| = 2.61+0.15-0.16 ({\rm stat.}) \pm 0.09 ({\rm syst.}) (\times 10^{-3} {\rm eV^2})$. The 5 MeV excess dependency on the reactor thermal power rate is again clearly observed with the increased data set.}

\FullConference{The 19th International Workshop on Neutrinos from Accelerators-NUFACT2017\\
		25-30 September, 2017\\
		Uppsala University, Uppsala, Sweden}

\begin{document}

\section{Introduction}
RENO has successully measured the value of the smallest neutrino mixing angle, $\theta_{13}$, and has
undertaken the measurement of the squared mass difference $|\Delta m^2 _{ee}|$.
The inverse beta decay (IBD) data collected at RENO uses electron antineutrinos
produced by six equally spaced reactors of the Hanbit nuclear power plant. There are two identical detectors located at near and
far sites at 294 m and 1,383 m, respectively, from the center of the reactor
array. The power plant consists of six equally spaced reactor cores placed
linearly and provides a total thermal power of 16.8 GW$_{\rm th}$ in full operation mode.

Even though all three neutrino mixing angles and two mass square
differences in the PMNS matrix are measured based on reactor experiments, precise measurements of these parameters
are still important issues for current and future neutrino oscillation experiments to measure leptonic CP
violation and to determine neutrino mass ordering. Recently some methodologies to measure neutrino mass ordering using reactor anti electron neutrino are discussed \cite{1}.

In this work RENO improved background
systematic uncertainty and updated $\sin^2 2\theta_{13}$ and $|\Delta m^2 _{ee}|$
using 1,500 live days of data collected in the
detectors. The updated 5 MeV excess estimation is also reported.

\section{RENO Detector}
RENO near (far) detector is constructed with 120 (450) m.w.e. overburden. The two detectors
are assembled identically in a concentric cylindrical shape. Each detector consists of inner detector (ID) and outer veto
detector (OD) filled with 350 ton purified water. The ID consists of target (16 ton liquid scintillator
with 0.1\% Gd), $\gamma$-catcher (29 ton liquid scintillator), and buffer (65 ton mineral oil) from the detector center. Total 354 (67)
Hamamatsu 10 inch PMTs are installed on the buffer (veto) wall. More details on the RENO experimental setup and the
detector are found in \cite{2,3}.

\section{Data Sample}
RENO has been taking data since 2011 continuously with average DAQ live time
efficiency of ~95\% for both detectors. In this analysis we use data collected from Aug. 19, 2011 to Apr.
23, 2017 for near detector and from Aug. 11, 2011 to Sep. 23, 2015 for far detector. Total live time of
the data is 15,47.35 (1,397.78) days for near (far) detector.

\begin{table}
\center
\begin{tabular}{c c c }
\hline
Detector & Near & Far \\ \hline
Selected candidate events & 732.168&68.055\\
Total background rate & 9.34$\pm$0.37&1.95$\pm$0.15\\
IBD rate after background subtraction&463.80$\pm$0.66&46.75$\pm$0.24\\ \hline \hline
Livetime [days]& 1,547.35&1,397.78\\ \hline
Accidental & 2.07$\pm$0.02&0.38$\pm$0.01\\
$^{9}$Li/$^{8}$He & 5.49$\pm$0.36&0.93$\pm$0.15\\
Fast neutron&1.74$\pm$0.02&0.35$\pm$0.01\\
$^{252}$Cf&0.04$\pm$0.01&0.28$\pm$0.02\\ \hline
\end{tabular}
\caption{Event rates per day of the observed IBD candidates and the estimated background in 1.2 < E$_p$ < 8 MeV}
%\end{ruledtabular}
\end{table}

RENO select IBD event sample by applying the IBD selection criteria described in \cite{2}. In this analysis, to reduce
background rate and its uncertainty, the optimized values of the spatial coincidence
requirement of $\Delta$R < 2.0 m to lower the accidental background is considered. The following multiplicity
requirements are also changed to make additional reduction of fast neutron, $^9$Li/$^{8}$He and $^{252}$Cf
backgrounds (note that the indexes of changed criteria are the ones used in \cite{3}): a timing veto
requirement for rejecting coincidence pair (a) if they are accompanied by any preceding ID or OD
trigger within a 300 $\mu$s window before their prompt candidate, (b) if they are followed by any
subsequent ID-only trigger other than those associated with the delayed candidate within a 200 (800)
$\mu$s window from their prompt candidate (only far $^{252}$Cf contaminated data), (d) if there are other
subsequent pairs within the 500 (1,000) $\mu$s interval (only far $^{252}$Cf contaminated data), (f) if they are
accompanied by a prompt candidate of E$_P$ > 3 MeV and Q$_{max}$/Q$_{tot}$ < 0.04 within a 10 (20) s window
and a distance of 40 (50) cm for near (far) $^{252}$Cf contaminated data; (ix) a spatial veto requirement for
rejecting coincidence pairs in the far detector only if the vertices of their prompt candidates are located
in a cylindrical volume of 50 cm in radius, centered at x = +12.5 cm and y = +12.5 cm and z < -110 cm. Total dead time due to the selection criteria is estimated as 40.37 (31.47)\% for near (far)
data. The same detection efficiency in \cite{4} is used in this analysis.

Some background could remain in the IBD candidate events sample passing the selection criteria.
The methods to estimate the remaining background are well described in \cite{4} and adopting the same
method for the 1,500 live days of RENO data we estimated the remaining background and summarized
in Table 1.
\begin{table}
\center
\begin{tabular}{c c c }
\hline
 & Bin-correlated & Bin-uncorrelated \\ \hline
Total background & 0.60\% (near), 1.99\% (far)&3.94\% (near), 2.71\% (far)\\ \hline
Accidental & 0.37\% (near), 0.96\% (far)&0.18\% (near), 0.49\% (far)\\
$^{9}$Li/$^{8}$He &1.01\% (near), 3.66\% (far)&6.71\% (near), 4.17\% (far)\\
Fast neutron&0.23\% (near), 0.54\% (far)&0.75\% (near), 0.83\% (far)\\
$^{252}$Cf&6.00\% (near), 1.11\% (far)&10.23\% (near), 12.62\% (far)\\ \hline
\end{tabular}
\caption{Background systematic uncertainties in 1.2 < E$_p$ < 8 MeV}
%\end{ruledtabular}
\end{table}

\section{Systematic Uncertainties}
To obtain the systematic uncertainties, the methods described in \cite{4} is applied. Based on these
methods we estimated our systematic uncertainties on background and summarized in Table 2. The
systematic uncertainties of reactor, detection efficiency including timing veto, and energy scale remain
the same as before \cite{4}.

\begin{figure}[htbp]
\centering
\includegraphics[width=120mm]{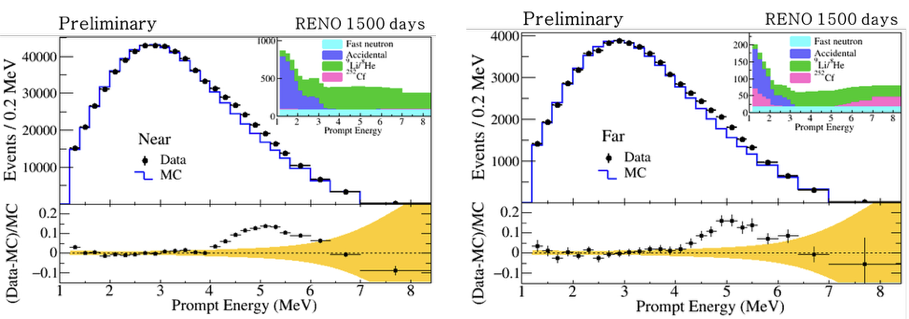}
\caption{\label{fig1}(Top panels) Comparison of the IBD prompt spectra between 1,500 live days of RENO data
and expectation \cite{6,7}. (Bottom panels) The fractional difference between the observed and expected
spectra where the 5 MeV excesses are clearly shown in both near and far data.}
\end{figure}

\begin{figure}[htbp]
\centering
\includegraphics[width=120mm]{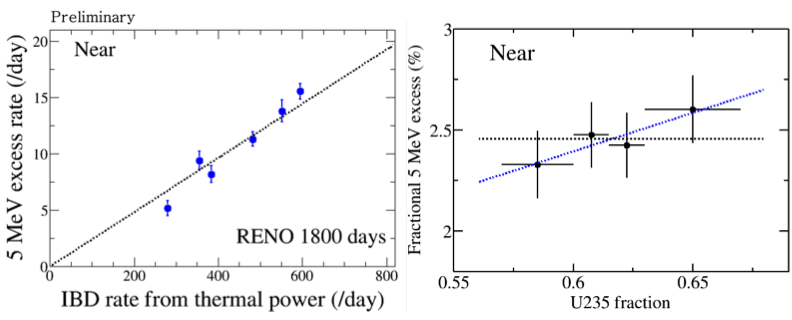}
\caption{\label{fig1}((Left) The 5 MeV excess vs. IBD rates per day. (Right) The 5 MeV excess vs. $^{235}$U fission
fraction.}
\end{figure}

\section{Results}
With 1,500 live days of data we estimate the 5 MeV excess and spectral measurement of
$\sin^2 2 \theta_{13}$ and $|\Delta m^2 _{ee}|$.

\begin{figure}
\centering
\includegraphics[width=80mm]{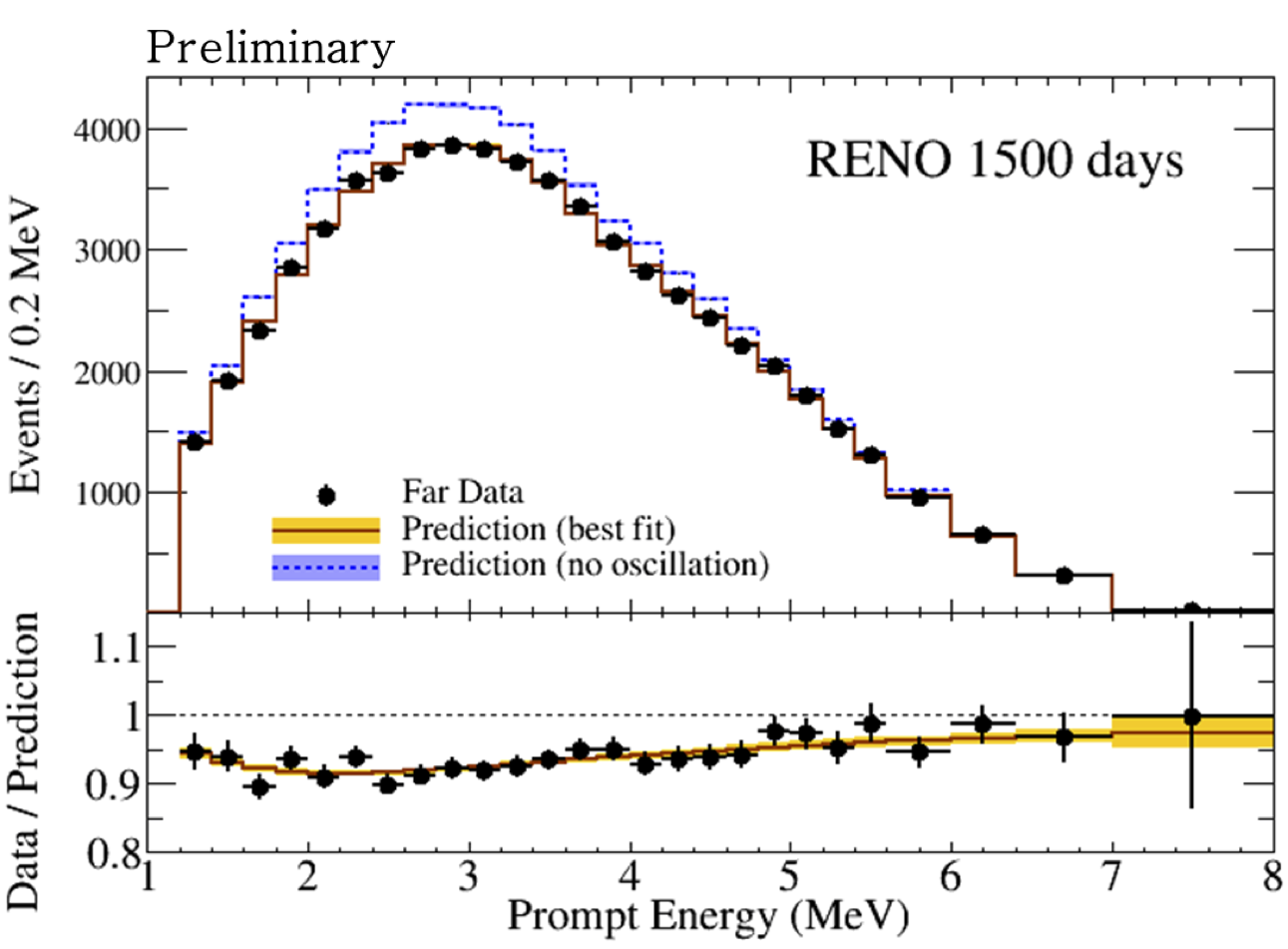}
\caption{(Top panel) Observed (black dots with
error bars) vs. expected (blue dotted histogram)
IBD prompt energy spectra after background
subtraction at far site. The expected spectrum at
far site is obtained using the near IBD data
assuming no oscillation. The orange histogram
represents the expected IBD spectrum with bestfit
oscillation parameters. (Bottom panel) Ratio
of the observed to the expected IBD prompt
spectra. There is a clear energy dependent
reactor neutrino disappearance.}
\end{figure}

\begin{figure}
\centering
\includegraphics[width=80mm]{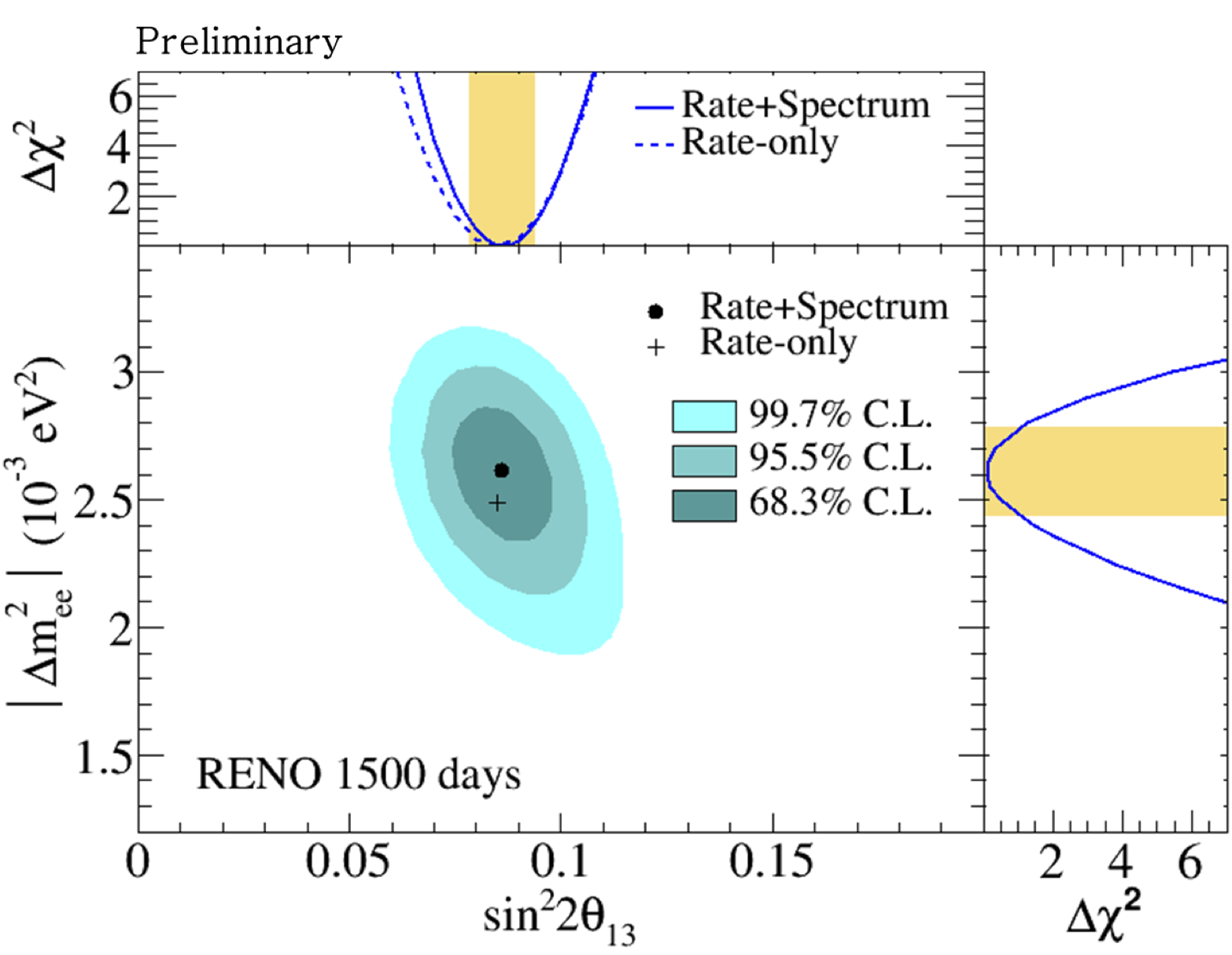}
\caption{(Contour plot of $\sin^2 2 \theta_{13}$ vs. $|\Delta m^2 _{ee}|$. The
best fit value for rate + shape (rate-only assuming
$|\Delta m^2 _{ee}| = 2.49 \times 10^{-3} {\rm eV}^2$) analysis is represented as
a black dot (cross). The three ellipses represent
the corresponding confidence levels of 68.3\%,
95.5\%, and 99.7\%. The upper (righter) panel
shows 1-dimentional $\Delta \chi^2$ distribution for $\sin^2 2 \theta_{13}$
($|\Delta m^2 _{ee}|$) and 1 $\sigma$ error band in orange color.}

\end{figure}

RENO is the first reactor neutrino experiment group who discovers the 5 MeV excess based on spectral comparion of observed and expected IBD prompt events at the two detectors in 2014 using 800 live days of RENO data \cite{5}. The
correlation between the 5 MeV excess and the IBD rate, i.e. the reactor thermal power was also
reported. These results are updated using 1,500 live days of data. Figure 1 top panels show observed
IBD prompt spectra of near and far data compared to the expected ones by the Huber and Mueller
model \cite{6,7} normalized to the area except the 5 MeV excess region. 
The bottom panels of the Fig. 1
depict the difference between the two spectra in the corresponding upper panels, where yellow bands
represent uncertainties in the model. A clear spectral discrepancy is observed
in the region of 5 MeV in both detectors. For the spectral
comparison only, the MC predicted energy spectra are
normalized to the observed events out of the excess range
3.6 < E$_{p}$ < 6.6 MeV. The excess of events is estimated as about 2.5\% of the total observed reactor 
$\bar{\nu}_{e}$ rate in both
detectors. 

\begin{figure}
\centering
\includegraphics[width=80mm]{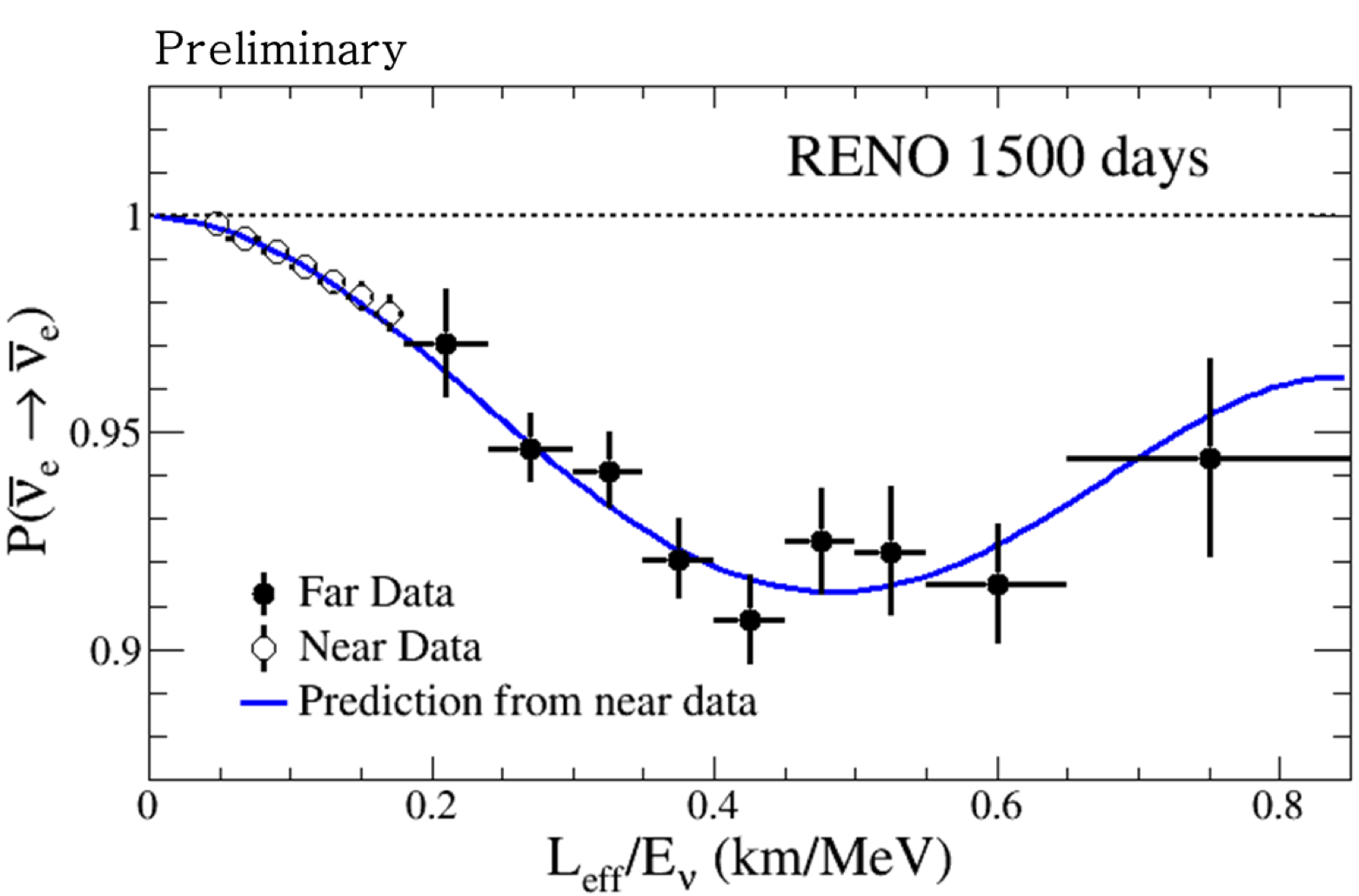}
\caption{Reactor neutrino survival
probability as a function of L$_{\rm eff}$/E. The
L$_{\rm eff}$ is a flux weighted effective distance
to a detector from the six reactors with
different baselines. Far data matches
well with the best-fit oscillation
prediction (blue curve).}
\end{figure}

Using the $\chi^2$ function for the rate + shape analysis described in \cite{3}, $\sin^2 2\theta_{13}$ and $|\Delta m^2 _{ee}| $ are obtained for
the 1,500 live days of RENO data. The measured values using events in 1.2 < E$_{p}$ < 8 MeV are:
$\sin^2 2\theta_{13} = 0.086 \pm 0.006 ({\rm stat.}) ± 0.005 ({\rm syst.})$ and 
$|\Delta m^2 _{ee}| = 2.61+0.15-0.16 ({\rm stat.}) \pm 0.09 ({\rm syst.}) (\times 10^{-3} {\rm eV^2})$.
The total uncertainty on $\sin^2 2\theta_{13}$ ($|\Delta m^2 _{ee}| $) is reduced from 12 (10)\% to 9 (7)\% compared to our
previous measurements using 500 live days of data \cite{3}. Figure 3 top panel shows the observed IBD
prompt spectrum at far (black dots with error bars) and the expected one obtained from near data
assuming no oscillation. There is a clear discrepancy between the two due to electron anti neutrino
disappearance at far, and their ratio is drawn in the bottom panel where the energy dependent
discrepancy is shown well. Figure 4 shows the contour plot and the best-fit values of rate + shape
(black dot) and rate-only (cross sign) measurements. Figure 5 shows the electron anti neutrino survival
probability as a function of L$_{\rm eff}$/E. Both near (open circles) and far (black dots) data points are shown
with the best-fit oscillation probability (blue curve). The far data points matches very well to the best-fit
oscillation. The near data points, however, matches extremely well to the best-fit oscillation since
the near expectation without oscillation was obtained by unavoidably using near data itself rather than
MC. Note that MC can not be used in this case because of the mismatch in the 5 MeV excess region.

In summary, using 1,500 live days of data RENO has reduced the uncertainties to 9\% and 7\% for the $\sin^2 2\theta_{13}$ and
$|\Delta m^2 _{ee}| $measurements, respectively. RENO has a plan to  reduce the $\sin^2 2\theta_{13}$ uncertainty to ~6\% using
data taken by 2018. With additional 2 or 3 more years of data taking from 2019 the uncertainty on the
$|\Delta m^2 _{ee}| $ measurement is expected to be reduced to 4$\sim$5\% even though the $\sin^2 2\theta_{13}$ uncertainty would
remain as ~6\%.

\end{document}